\documentstyle{article}

\newcommand{\nc}{\newcommand}
\nc{\bea}{\begin{eqnarray}}
\nc{\eea}{\end{eqnarray}}
\nc{\beq}{\begin{equation}}
\nc{\eeq}{\end{equation}}
\nc{\nn}{\nonumber}
\nc{\st}{\scriptstyle}

\def\/{\over}
\begin{document}

\parindent=1 em
\frenchspacing

\author{
\large  Rainer M\"uller and Carlos O. Lousto  \\
\normalsize \it Fakult\"at f\"ur Physik der Universit\"at Konstanz\\
\normalsize \it Postfach 5560 M 674, D-78434 Konstanz, Germany}

\title{
Recovery of information from black hole radiation by considering
stimulated emission}
\date{}

\maketitle
\vspace{5mm}

\begin{abstract}
We deal with the black hole
information loss paradox by showing that the stimulated
emission component of the black hole radiation
contains information about the initial state of the
system. The nonlocal
behaviour that allows the recovery of information about the
matter that falls behind the horizon appears in a natural
way. We calculate the  expectation value and probability
distribution of particles at ${\cal J}^+$ for a
non-vacuum initial state. The entropy
of the final state is compared to that of a thermal state
with the same energy per mode. We find that the information
recovered about the initial state increases with the number
$r$ of the initially incoming particles, reaching for example
over 30\% for $r = 1000$.

We point out that recovering information about the initial
state, nevertheless, does not automatically imply
the purity of the final state.
\end{abstract}
\vspace{1cm}

\section{Introduction}

Currently there is a vivid discussion about the so-called
information-loss paradox that appears when
a matter distribution collapses to form a black hole and
subsequently evaporates into Hawking
radiation. Consider a matter distribution that is
described quantum mechanically by a pure
state (zero entropy) and collapses, forming a black hole.
The black hole will emit Hawking radiation
and, taking gravitational back-reaction into account, is
believed to evaporate almost completely.
After the evaporation of the black hole, only the thermal
Hawking radiation is left behind, which
is described by a density matrix. Hence it appears as if
the system had evolved from an initial
pure state to a final mixed state so that information must
have been lost.

Let us examine what kind of information has been lost.
There are two sources of
information loss, which may be related, but which are not
always properly distinguished in the literature: (1) After
the complete black hole evaporation the
total system consists solely of Hawking radiation. Thermal
radiation cannot carry any information except for
its temperature. It depends only on the geomety of
the black hole and does not
reveal anything about the matter that formed the black hole
(no hair). Thus, even if the
initial state of the collapsing matter that fell behind the
horizon was precisely known,
this information will not be present in the thermal
Hawking radiation of the final state.
(2) When the Hawking radiation is created, the total system
consists of two subsystems:
The interior and the exterior region of the black hole. In
such cases it is known
\cite{Zurek83,Barnett89} that, although the state of the
total system is pure, information
can be transferred from each of the subsystems into
correlations between the subsystems.
The state of each subsystem is then described by a density
matrix. This is also the case for
a black hole and the radiation it emits. When the black
hole evaporates, it is not clear what happens to the
information contained in correlations.

There have been several attempts to resolve this
information loss paradox. Each of them is
plagued of serious objections, however. Review of the
subject are given in Refs.
\cite{Preskill92,Page93,Page92,Wald92,tHooft92,Giddings92,Wald84}.
There are three main approaches to the problem:

The first possibility is to assume that the black hole does
not evaporate completely but
leaves behind a Planck mass remnant \cite{Aharonov87}.
This proposal has problems with the
infinite number of remnant species and their specific
entropy and is therefore disregarded
by most authors.

A second alternative, advocated by Hawking \cite{Hawking76},
is to accept the information loss
and the evolution from pure into mixed states. This would
mean that the usual rules of
quantum mechanics would not apply to the problem of black
hole evaporation
(in particular there are problems with locality and energy
conservation \cite{Banks84}). Unitary evolution seems
to be lost at least at the level of quantum gravity.

The third possibility is that the information is not lost
and the final state is pure. This has been
put forward mainly by Page \cite{Page80,Page92} and
't Hooft \cite{tHooft92}.
In this case, the information should be completely encoded
in the Hawking radiation.
One is then faced with the problem that the information
must be extracted from behind the
horizon in a nonlocal manner \cite{Wald86}. Up to now, no
explicit mechanism has been given for this nonlocal
process.

In the present paper we study a mechanism that encodes
information
about the initial state into the Hawking radiation. We
found already in a previous paper
\cite{Audretsch92} that the outgoing Hawking radiation can
be influenced by ingoing
particles via stimulated emission, even if they fall behind
the horizon. The imprints left by the stimulated emission on
the black hole radiation carry information about the particles
that formed the black hole. They can, in principle, be
measured by an external observer that could thus estimate
the initial state of the system. It is remarkable that the
nonlocal behaviour
that is necessary for obtaining information about the
matter that crosses the horizon appears in the stimulated
emission process in a natural way.

Since our calculation is valid only near the time of
horizon formation, we are not able to treat
infalling matter at late times. The role of stimulated
emission in this situation has been recently
invesigated by Bekenstein \cite{Bekenstein93} and Schiffer
\cite{Schiffer93}.

We will show, however, that although the stimulated
emission process provides a mechanism for information
extraction, this does not necessarily imply the purity of the
final state. The problems of recovering
information and obtaining a pure state must therefore be
carefully distinguished.

In Sec.\ 2, we review the main results of the computation
of the stimulated emission component of the black
hole radiation. We then compute in Sec.\ 3 the entropy associated
with this radiation and compare it with the entropy of a thermal
state with the same mean energy to have a measure of
the information carried by the Hawking radiation. We find
that, for example, for 100 incoming particles per mode as much as
28\% of the information can be recovered. The paper ends
with a discussion of the results and the Appendix where we
discuss explicitly the nonlocal terms of the stimulated emission.

\section{Modification of Hawking radiation by
stimulated emission}

We consider the spacetime generated by a spherically
symmetric matter distribution that collapses to form
a black hole (Fig. 1). Outside the matter the metric is the
Schwarzschild metric which can be covered by the
Eddington-Finkelstein null  coordinates \cite{Misner73}
$u=t-r^\ast$ and $v=t+r^\ast$
where  $r^\ast =r+2M\ln\left|{r\/ 2M}-1\right|$. Radially ingoing
null
geodesics $v=\hbox{const}$ and outgoing null geodesics
$u=\hbox{const}$ are connected
inside the matter distribution by Hawking's ray tracing
formula \cite{Hawking75}
\beq v = v_0 - C D \, e^{-\kappa u},  \label{eq1} \eeq
where $C$, $D$ are constants and $\kappa = (4M)^{-1}$ is
the black hole's surface
gravity. The ingoing ray that reaches the horizon at the
moment of its formation is denoted by
$v_0$. Eq. (\ref{eq1}) is valid for $v-v_0$ small
and positive. Therefore we
restrict our considerations to late times (large $u$) at
${\cal J}^+$. It is possible to derive the
redshift suffered by a null ray that follows the trajectory
(\ref{eq1}) in the limit of geometrical
optics. It is given by the redshift formula
\cite{Audretsch92}
\beq  \omega = \kappa C D \,e^{-\kappa u}\, \omega'.
\label{eq2}  \eeq
Throughout the paper, primed quantities refer to
observables at ${\cal J}^-$, unprimed ones to ${\cal J}^+$.

Let us consider a real massless scalar field in this
spacetime. In order to quantize the field
we need a complete set of mode functions. We choose ingoing
and outgoing wave
packets which are positive frequency on ${\cal J}^-$ and
${\cal J}^+$, respectively.
\beq f_{\omega' n' l' m'} =
{1\/\sqrt{\epsilon'}}\int_{\omega'}^{\omega'+\epsilon'}
d{\tilde \omega'}
e^{-i {\tilde \omega'} n'} f_{{\tilde \omega'} l' m'} ,
\qquad
 p_{\omega n l m} =
{1\/\sqrt{\epsilon}}\int_{\omega}^{\omega+\epsilon}
d{\tilde \omega} \,
e^{-i {\tilde \omega} n} f_{{\tilde \omega} l m}
\label{eq3} \eeq
where $\omega' / \epsilon'$, $\omega / \epsilon$,
$n'\epsilon' / 2\pi$ and $n\epsilon / 2\pi$
have to be integers. They are formed from the spherical
waves \cite{Hawking75}
\beq f_{\omega' l' m'} = N \omega'^{-{1\/2}} e^{-i\omega'
v}\,{1\/ r}Y_{l'm'} , \qquad
   p_{\omega l m} = N \omega^{-{1\/2}}e^{-i\omega u}\,{1\/
   r}Y_{lm}, \label{eq4} \eeq
whose angular dependence is given by the spherical
harmonics $Y_{lm}$. $N$ is a normalization
factor. They are solutions of the Klein-Gordon equation in
the Schwarzschild geometry if gravitational
backscattering is neglected. The parameters $(\omega' ,n')$
and $(\omega , n)$ of the
wave packets (\ref{eq3}) specify their mean energy and
their trajectory according to
\beq  v = -n', \qquad u=-n.  \label{eq5} \eeq
With trajectory we refer to the path of the wave packet
maximum. One must not forget, however, that a wave packet
inevitably has tails that extend over the whole spacetime.

The quantization is carried out with respect to the wave
packet mode functions
(\ref{eq3}) on ${\cal J}^-$
\beq \phi (x) = \sum_{\omega' n' l' m'} \left(
b^{in}_{\omega' n'l'm'} f_{\omega' n'l'm'}(x) +
	b^{in\dagger}_{\omega' n'l'm'}
f^{\ast}_{\omega'n'l'm'}(x) \right) , \label{eq6}   \eeq
and on  ${\cal J}^+$
\beq \phi (x) = \sum_{\omega n lm} \left( a^{out}_{\omega
nlm} p_{\omega nlm}(x) +
	a^{out\dagger}_{\omega nlm} p^{\ast}_{\omega n
	lm}(x) \right) , \label{eq7}   \eeq
resulting in wave packet creation
and annihilation operators.
To describe particles at the future event horizon $H^+$,
we choose wave packets formed from Wald's horizon modes
\cite{Wald75}. The choice of these mode functions does
not affect observables at ${\cal J}^+$.

In Ref. \cite{Audretsch92}, we have obtained the Bogoliubov
transformation that connects creation
and annihilation operators on ${\cal J}^-$ and ${\cal
J}^+$. It can be written as
\beq a^{out}_{\omega nlm} = \alpha^\ast_{\omega lm}
b^{in}_{\omega'_\alpha n'_\alpha lm}
	-  \beta^\ast_{\omega lm} b^{in\dagger}_{\omega'_\beta
	n'_\beta l,-m} .  \label{eq8}\eeq
The approximations contained in the derivation of
(\ref{eq8}) are discussed in \cite{Audretsch92}.
In (\ref{eq8}) we introduced the notion of the equivalent
mode \cite{Audretsch93}
of a wave packet $(\omega, n, l, m)$ at ${\cal J}^+$:
\beq ( \omega'_\alpha, n'_\alpha, l, m ) =
\left(  {1\/ \kappa
C D} e^{-\kappa n} \omega, \quad
	-( v_0 - C D \, e^{\kappa n}), l, m  \right).
\label{eq9}  \eeq
It is given by the wave packet $ ( \omega'_\alpha,
n'_\alpha, l, m )$ at ${\cal J}^-$ which is connected
to $(\omega ,n, l, m)$ by means of the classical trajectory
(\ref{eq1}) and redshift relation (\ref{eq2}) (cf. Fig.\
1).The mirror mode
\beq ( \omega'_\beta, n'_\beta, l, -m ) = \left(  {1\/
\kappa C D} e^{-\kappa n} \omega, \quad
	-( v_0 + C D \, e^{\kappa n}), l, -m  \right)
\label{eq10}  \eeq
of a wave packet $(\omega, n, l, m)$ which enters the
horizon is obtained from the equivalent
mode by `reflection' at the null line $v_0$. Furthermore
we used in (\ref{eq8}) the Bogoliubov
parameters
\beq\left. {\alpha_{\omega lm} \atop \beta_{\omega lm}}
\right\} =
	\mp i \left( {\kappa \/2\pi \omega} \right)^{1\/2}
	e^{\pm \omega \pi / 2 \kappa} \,
	\left| \Gamma\left( 1 + i {\omega \/\kappa} \right)
	\right|   (-1)^l \, (\pm 1)^m.
\label{eq11}  \eeq
Their squares are given by
\beq |\alpha_{\omega lm}|^2  = 1 +  {1\/
  e^{\omega/T}-1},   \qquad
   |\beta_{\omega lm}|^2  = {1\/ e^{\omega/T}-1},
 \label{eq12}  \eeq
where $T= \kappa / (2 \pi)$ is the black hole temperature.

With these results, it is possible to discuss the
spontaneous and stimulated emission of the
Hawking radiation. To this end, we calculate the
expectation value of the outgoing particle
number operator, $r^{out}_{\omega nlm} = \allowbreak
\langle\psi^{in} | a^{out\dagger}_{\omega nlm}
a^{out}_{\omega nlm} | \psi^{in}\rangle $, that is measured
at ${\cal J}^+$ for a given initial
state $\psi^{in}$ prepared at ${\cal J}^-$. We find
\beq r^{out}_{\omega nlm} = r^{in}_{\omega'_\alpha
n'_\alpha lm} + |\beta_{\omega lm}|^2
	( r^{in}_{\omega'_\alpha n'_\alpha lm} +
	r^{in}_{\omega'_\beta n'_\beta l,-m})
	+ |\beta_{\omega lm}|^2.   \label{eq13} \eeq
Here $r^{in}_{\omega' n'l'm'}= \langle\psi^{in} |
b^{in\dagger}_{\omega' n'l'm'}
b^{in}_{\omega' n'l'm'} |\psi^{in}\rangle $ denotes the
mean number of ingoing particles in the state
$| \psi^{in}\rangle $
at ${\cal J}^-$. Eq.\ (\ref{eq13}) shows how the number
of particles in the black hole radiation depends on the
particle number in the initial state.

We see that there are four contributions to the particle
number at ${\cal J}^+$. The last term
is the thermal spectrum of the Hawking radiation, which is
always present, regardless
of the of the initial state of the field. The remaining
terms show that in addition to this thermal
radiation, there are further contributions due to the
presence of ingoing particles.
The first term represents the particles that escaped the
capture by the horizon $(v<v_0)$
and reappear in the out region in the equivalent mode
(\ref{eq9}). They travel on the
classical null geodesics (\ref{eq1}) with the respective
redshift (\ref{eq2}).  The second term
shows that these particles stimulate the creation of
additional particles on the same trajectory
with the same energy. Finally, we see from the third term
that the particle number in a wave
packet on ${\cal J}^+$ can also be enhanced by ingoing
wave packets
that travel {\it behind} the horizon $(v>v_0)$ in the
mirror mode (\ref{eq10}). In the
Appendix, we show the explicit nonlocality of this
contribution.

It is possible to give a heuristic discussion of the
physics contained in Eq. (\ref{eq13}):
The black hole radiation is created in
pairs. The members of each pair
are located on either side of the horizon
\cite{Wald75,Hawking76}. Ingoing
wave packets that reach ${\cal J}^+$ stimulate the creation
of additional pairs in the same
mode. The additional particles appear as the second term in
(\ref{eq13}). But the creation
of additional pairs can also be stimulated by
wave packets that travel
behind the horizon in the mirror mode (\ref{eq10}). The
respective particles then reach
${\cal J}^+$, leading to the nonlocal term in
(\ref{eq13}).

\section{Entropy of the black hole radiation}

Let us calculate the entropy of the black hole radiation
when stimulated emission is taken into account. This
can be compared with the entropy of thermal radiation with
the same mean energy. The deviation of the black hole
radiation from thermality is a measure for the amount of
information that can be encoded in the
black hole radiation.

To this end we transcribe an incoming many-particle state
in terms of wave packet states at ${\cal J}^+$ and
the horizon $H^+$. This
procedure has been carried out for a Bogoliubov
transformation of the form (\ref{eq8}) in the context of
accelerated observers in Ref.\ \cite{Audretsch92}
(cf. also \cite{Gasperini90}).We therefore give here
only the relevant results. Consider
a wave packet state $| r^{in}_{\omega'_\beta n'_\beta l,-m}
\rangle$ with $r$ ingoing particles that cross the
event horizon. With respect to particles at ${\cal J}^+$
and $H^+$, this state can be written
\beq
| r^{in}_{\omega'_\beta n'_\beta l,-m} \rangle =
  | r_{\omega n l m} \rangle
  \prod_{{\tilde \omega \tilde n \tilde l \tilde m} \atop
  {\neq \omega n l m}} | 0_{\tilde \omega \tilde n
  \tilde l \tilde m}\rangle,
\label{eq21} \eeq
where $(\omega, n, l, m)$ and $(\omega'_\beta, n'_\beta,
l, -m)$
are connected according to (\ref{eq10}) and the product
extends over all remaining modes. Furthermore we have
defined
\beq
| r_{\omega n l m} \rangle = {(-\alpha_{\omega lm})^{-r}
  \/ | \alpha_{\omega lm}| } \sum_{q=0}^\infty
  \left({(r+q)! \/ r! q!}\right)^{1\/ 2} \left(
  \beta^\ast_{\omega l m}
  \/ \alpha_{\omega l m} \right)^q |(r+q)_{\omega nlm}
  \rangle_{H^+} \otimes | q_{\omega nlm} \rangle_{\cal
  J^+}.     \label{eq22} \eeq
The state (\ref{eq21}) displays nonlocal correlations
between particles at ${\cal J}^+$ and $H^+$.

The state (\ref{eq21}) is pure. To calculate the entropy
of the black hole radiation at ${\cal J}^+$ we need the
reduced density matrix which is obtained by tracing
over the horizon states. Since in our approximation ingoing
particles in the
mode $(\omega'_\beta, n'_\beta, l,-m)$ influence only
the mode $(\omega, n, l, m)$,
we restrict our attention to this mode. Its reduced density
matrix is
\bea
\rho^{\cal J^+}_{\omega n l m} &=& \hbox{Tr}_{H^+}
  (|r_{\omega nlm} \rangle \langle r_{\omega nlm} |)
  \nonumber \\
  &=& |\alpha_{\omega nlm}|^{-2r-2} \sum_{q=0}^\infty
  {(r+q)! \/ r! q!} \left| \beta_{\omega lm} \/
  \alpha_{\omega lm} \right|^{2q} |q_{\omega nlm}
  \rangle_{\cal J^+ J^+}\langle q_{\omega nlm}|.
\label{eq23} \eea
The entropy of the black hole radiation in the mode
$(\omega, n, l, m)$ at ${\cal J}^+$ is given by
\bea
S^{bh}_{\omega nlm} &=& -k \,\hbox{Tr}_{\cal J^+}
  (\rho^{\cal J^+}_{\omega n l m} \ln
  \rho^{\cal J^+}_{\omega n l m}) \nonumber \\
  &=& - k \,|\alpha_{\omega lm}|^{-2r-2}
  \sum_{q=0}^\infty {(r+q)!\/ r! q!} \left| \beta_{\omega
  lm} \/ \alpha_{\omega lm} \right|^{2q}
  \ln\left( |\alpha_{\omega lm}|^{-2r-2} {(r+q)!\/ r! q!}
  \left| \beta_{\omega lm} \/ \alpha_{\omega lm}
  \right|^{2q} \right) . \label{eq24} \eea
The sum can be evaluated using the formula
$$\sum_{q=0}^\infty {(r+q)!\/ r! q!} x^q =
  (1-x)^{-r-1}, $$
and the integral representation for the logarithm of the
gamma function \cite{Abramowitz}
\beq \ln \Gamma (z) = \int_0^\infty {dt\/t} \left(
  (z-1) e^{-t} + {e^{-zt}-e^{-t} \/ 1-e^{-t}} \right).
\eeq
We obtain
\beq S^{bh}_{\omega nlm} = k (r+1) \left( |\alpha_{\omega
  lm}|^2 \ln |\alpha_{\omega lm}|^2 - |\beta_{\omega lm}|^2
  \ln |\beta_{\omega lm}|^2 \right) + S^{bh}_R ,
\label{eq25} \eeq
where
\beq
S^{bh}_R = -k\, |\alpha_{\omega lm}|^{-2r-2} \int_0^\infty
  {dt\/t} {e^{-rt}-1 \/ e^t -1} \left[ \left(
  1- \left|\beta_{\omega lm}\/ \alpha_{\omega lm}\right|^2
  e^{-t}\right)^{-r-1} - \left( 1- \left|\beta_{\omega
  lm}\/ \alpha_{\omega lm}\right|^2 \right)^{-r-1}
  \right] .\label{eq26} \eeq
The remaining integral has to be treated numerically.

This expression for the  entropy contained in the mode
$(\omega, n, l, m)$ of
the black  hole radiation in the presence of incoming
particles
can be compared with the entropy $S^{th}_{\omega nlm}$
of thermal radiation in this mode with the particle
number expectation value $ r^{th}_{\omega nlm}
 = |\beta_{\omega lm}|^2 ( r^{in}_{\omega'_\beta n'_\beta
l,-m}+1) $ obtained from (\ref{eq13}). It is given by
\cite{Yamamoto86}
\beq S^{th}_{\omega nlm} = k \left( (1 + r^{th}_{\omega
nlm}) \ln ( 1 + r^{th}_{\omega nlm}) - r^{th}_{\omega nlm}
\ln r^{th}_{\omega nlm} \right). \label{eq28} \eeq

We can take $I_{\omega n l m} = (S^{th}_{\omega nlm}
- S^{bh}_{\omega nlm})/ S^{th}_{\omega nlm}$
as a measure of the amount of information contained
in the wave packet $(\omega,n,l,m)$ at ${\cal J}^+$
about the initial mode $(\omega_\beta, n_\beta, l,-m)$
that propagates behind the horizon. It depends only on the
parameter $\omega /T$ and on the number $r$ of ingoing
particles. It is independent of the trajectory and angular
momentum of the respective wave packets. In Fig.\ 2 we
plotted $I_{\omega n l m}$ as a function of $|\beta_{
\omega l m}|^2$ for several values of $r$. Note that
generally the amount of recovered information increases
with the incident particle number $r$. In the region
$\omega > T$ (small $|\beta_{\omega l m}|^2$),
$I_{\omega n l m}$ increases with $|\beta_{\omega l m}|^2$,
reaching a maximum near $\omega \approx T$. For larger
$|\beta_{\omega l m}|^2$, $I_{\omega n l m}$ decreases
very slowly. For 1000 (100, 10) particles in the initial wave
packet, the maximal amount of recovered information is
35\% (28\%, 16\%).

In a more realistic calculation, we would have to take into
account the increasing temperature in the course of the
black hole evaporation. But for high temperatures,
as one can see by numerical studies,
$I_{\omega n l m}$ is almost independent of $T$, so that
there is only a small influence of the changing temperature
until the final stages of evaporation.

It may be worth noting that the same investigations can be
carried out for an incident wave packet $(\omega_\alpha,
n_\alpha, l,m)$ that does not enter the horizon
(cf. Fig. 1). In contrast to the behaviour of horizon-%
crossing modes, $I_{\omega n l m}$ is in this case close
to unity for small $|\beta_{\omega l m}|^2$. The wave
packet propagates nearly undisturbed to ${\cal J}^+$
so that almost all information it carries can be
recovered. For larger values of $|\beta_{\omega l m}|^2$,
$I_{\omega n l m}$ decreases because of the disturbances
that the black hole radiation introduces to the
semiclassical propagation.

Let us turn now to the second aspect of the information
loss paradox, the question about the information contained
in correlations. A quantitative measure of this information
can be defined by \cite{Barnett89}
\beq J_c = S^{bh}_{\cal J^+} + S^{bh}_{H^+} -
  S^{bh}_{tot},   \label{eq29} \eeq
where $S^{bh}_{\cal J^+} = \sum_{\omega nlm} S^{bh}_{\omega
nlm}$ is the entropy of the black hole radiation at
${\cal J}^+$, $S^{bh}_{H^+}$ is the entropy at the horizon,
and $S^{bh}_{tot}$ is the entropy of the total state
(\ref{eq21}) which is zero in our case. Using the
Araki-Lieb inequalities \cite{Araki70}, this can be reduced
to
\beq J_c = 2 \sum_{\omega nlm} S^{bh}_{\omega nlm}.
  \label{eq30} \eeq
This expression for the information contained in
correlations in the case of black hole radiation can again
be compared with the same quantity for thermal radiation
which can be obtained by replacing in (\ref{eq30})
$S^{bh}_{\omega nlm}$ by $S^{th}_{\omega nlm}$ of
(\ref{eq28}). We see that the information contained in
correlations is smaller for black hole radiation than
for thermal radiation. Note however, that a
pure state for the black hole radiation would require
$J_c =0 $, which cannot be achieved by considering
the stimulated emission process alone.

It is possible to compute not
only the expectation value of the particle number,
but also the probability distribution of particles
at ${\cal J}^+$ in the presence of incoming
particles at ${\cal J}^-$. This quantity can be easily
derived from (\ref{eq21}).
The probability that $q$ particles are found in the mode
$(\omega, n,l,m)$ at ${\cal J}^+$
if the initial state is the vacuum is the well-known
Bose-Einstein distribution
(cf. \cite{Parker75})
\bea
P(q^{out}_{\omega n l m} | 0^{in} ) &=&
	{|\beta_{\omega lm}|^{2q} \/ (1+|\beta_{\omega
	lm}|^2)^{q+1}}   \nonumber \\
	& =&  \left(1-e^{-2\pi\omega/\kappa}\right)
	e^{-(2\pi\omega/\kappa)q} , \label{eq14}
\eea
which is independent of the trajectory and angular
momentum of the particles.

The situation changes however when there are ingoing
particles at  ${\cal J}^-$. Let us
consider the case when there are $r$ incoming
particles in the mode $( \omega'_\beta,
n'_\beta, l, -m)$ that propagates behind the horizon.
The probability for finding $q$ particles
in the mirror mode $(\omega, n, l ,m)$ at  ${\cal J}^+$
is then
\bea
P(q^{out}_{\omega n lm} | r^{in}_{\omega'_\beta n'_\beta
l,-m} ) &=&   {(r+q)!\/r!q!}
	{|\beta_{\omega lm}|^{2q}\/ (1+|\beta_{\omega
	lm}|^2)^{r+q+1}}  	\nonumber \\
	&=&  {(r+q)!\/r!q!}  e^{-(2\pi\omega/\kappa)q}
	\left(1-e^{-2\pi\omega/\kappa} \right)^{r+1} .
 \label{eq15} \eea
For the remaining modes, the probability distribution is
the same as that for the vacuum
(\ref{eq14}). The result (\ref{eq15}) shows that in
agreement with the previous discussion the
Hawking radiation at ${\cal J}^+$ can be influenced by
ingoing particles that propagate
behind the horizon. The key point is that the probability
distribution (\ref{eq15}), which refers
only to observables at ${\cal J}^+$, depends explicitly
on the number $r$ of ingoing particles.

It is possible to gain some further insight from (\ref{eq15}).
Let us first consider for fixed $q$ the maximum of the probability
distribution as a function of $|\beta_{\omega lm}|^2$. We find
$|\beta_{\omega lm}|^2_{max} = q/(r+1)$. This shows that as $r$
increases the maximum of the probability distribution is located
at lower values of $|\beta_{\omega lm}|^2$. This is in agreement
with the behaviour of $I_{\omega nlm}$ as shown in Fig. 2.
The value of (\ref{eq15}) at the maximum $|\beta_{\omega
lm}|^2_{max}$
can be computed and in the limit of large $r$ we obtain
${1\/q!} q^q e^{-q}$. This is independent of $r$ and shows
that for any $q$ the maximum of the probability distribution
does not change as $r\to \infty$. Thus there seems to exist a finite
limit for the amount of information that can be recovered
when the number of ingoing particles is increased.
This feature is also in agreement with our numerical studies.

\section{Discussion}

Let us examine to what extent the effect of stimulated
emission contributes to the resolution of the black hole
information loss
paradox. The first problem is the loss of information
about the initial state of the collapsing matter.
We have seen that when the effects of stimulated emission
are taken into account, information concerning the
early stages, close to the
black hole formation, can be recovered in principle
at late times by measuring the
full Hawking radiation. In fact, we have shown that from
the knowledge of the probability distribution (\ref{eq15})
or the particle spectrum (\ref{eq13}), which refer only
to observables at ${\cal J}^+$, we can infer the initial
state at ${\cal J}^-$. Black hole radiation has a lower
entropy than thermal radiation with the same energy,
allowing information to come out. Thus, if
as is commonly said we form a black hole out of a pure
quantum state
$| \psi^{in} \rangle$, a part of the information it
carries can be recovered at late
times by measurement of the stimulated emission
contribution to the Hawking radiation.
Note that as always in quantum mechanics, one needs a
series of
measurements on identically prepared systems to obtain
a mean value or a probability ditribution.

We should mention that the energies of initial particles that
produce stimulated emission at late times will suffer a very
large redshift. This can be seen from Eqs. (\ref{eq2}) and
(\ref{eq10})
and has
been discussed previously in the literature
\cite{Wald76,Audretsch92}.
Accordingly, the stimulated emission contribution at late
times will have very low energies, making its detection difficult,
although the effect is present in principle.

The information loss paradox was put
forward \cite{Hawking76} by considering only the thermal
component of the Hawking
radiation (i.e. the last term in Eq. (\ref{eq13})).
However, to have a consistent
picture of the problem one must also consider the
contribution of the
stimulated emission. In fact, if one assumes that the
total mass of the
black hole formed from a pure state is a fraction
$\epsilon$ of the total energy content of the
initial state, i.e. $M=\sum_j\epsilon_j\,\omega_j^{in}$,
neglecting (as is
usually made) the effects of the stimulated emission would
mean
to neglect the contribution of the initial state, and thus
no black
hole that could emit Hawking radiation (thermal or not)
would in that case have formed.

In the previous sections, we restricted our investigations
to massless bosonic particles. To obtain generally
applicable results we must also consider massive particles
and fermions. For massive bosonic
particles the general structure of the stimulated emission
process
remains the same, although the details of particle creation
will be modified (cf. the general formulas in Ref.
\cite{Audretsch92}). For fermions, on the other hand, it
is well known that there is an
attenuation induced by ingoing particles instead of an
amplification \cite{Parker71,Audretsch86}.
It does also represent a deviation from thermality in
which information can be encoded.
This point has been discussed recently by Schiffer
\cite{Schiffer93}.

In a recent paper Bekenstein \cite{Bekenstein93} remarked
that by exploiting the
fact that the radiation emitted by a black hole is not
perfectly that of a black body, but distorted by a
barrier penetration factor $\Gamma(\omega)$,
information can leak out from the hole in the course of its
evaporation.
Indeed, additional contributions in Eq. (\ref{eq13}) can be
incorporated in a
$\Gamma(\omega)$ factor, and Bekenstein's considerations
would then
apply. However, stimulated emission is an important effect
in itself and
we consider it in an independent way. Also recently,
Schiffer \cite{Schiffer93}
has considered the effects of stimulated emission  for
bosons and fermions that fall in at late times, thus
making, in some sense, a
complementary study to ours.

Whereas the stimulated emission mechanism allows in
principle to recover information about the initial state
of the field, it does not solve the second part of the
information loss problem: The final state of the field at
${\cal J}^+$ is not pure because of our missing knowledge
about the detailed state of the thermal part of the black
hole radiation. The corresponding
information is contained in correlations.

We can compare the situation with the problem of signal
transmission through a noisy communication channel.
The initial state of the field plays the role of the
message, the thermal component of the black hole radiation
represents the noise. Even if  we could reconstruct
the complete message after receiving the noisy signal
(we are not interested in information about the noise),
this would not be tantamount to eliminating the noise.

In the black hole context, this means that
recovering information about the initial state does
not imply automatically the purity of the final state.
The problem is not the lack of information about the initial
state but about the final state of the system. Note the
double role played by  $|\beta_{\omega l m}|^2$: On one
hand, it is responsible for the noise in the final state
of the field (cf. (\ref{eq25})). On the other hand, the
recovery of information about the particles that crossed
the horizon is only possible for nonzero
 $|\beta_{\omega l m}|^2$ (cf. (\ref{eq13})). The two
aspects of  the paradox mentioned in the introduction
must therefore carefully be distinguished.

Probably, the definite answer to the information loss
paradox has to be given in the context of a full
quantum theory of gravitation. We can argue, however,
that the effects of stimulated emission will play an
important role in this resolution since, as we have
shown, they can carry a non-negligible part of the
information about the initial state of the system
(see Fig.\ 2).

In the light of the above results, there remain a
number of possibilities to solve the problem.
One alternative is
that the purity of the state is preserved at early
times. One could rely for example on the existence
of an environment that continuously `measures'
the system, thereby eliminating the lack of information
about the final state. Possibly,
the answer to the question has to be searched for
at  the late stages of evaporation \cite{Wilczek93}.
Perhaps a decrease in the number of degrees of freedom
of the evaporating black hole results in a reduced ability
to store information in correlations, thereby restoring
the purity of the state.
Finally there is the possibility that the nonunitarity appears
only in our present description of black hole evaporation.
This does not need to rule out the unitarity of quantum gravity,
however, since we are dealing with a semiclassical
theory. In fact, it has been shown recently
\cite{Kiefer91,Kiefer93} that nonunitary corrections appear if
the Wheeler-DeWitt equation is aproximated by an
effective equation for matter fields in a classical
background geometry. Thus the full quantum
gravitational description of black hole evaporation
may well be unitary.

\appendix

\section{Appendix: Angular distribution of the Hawking
radiation}

The wave packet mode functions (\ref{eq3}) are
eigenfunctions of angular momentum. Because
of their $Y_{lm}$ dependence, we were not able to obtain an
angular localization of the
Hawking radiation. Let us therefore introduce angular wave
packet modes at ${\cal J}^-$:
\beq  f_{\omega' n' \Theta' \Phi'} (v, \vartheta, \varphi)
= \sum_{l' m'} Y^\ast_{l' m'} (\Theta', \Phi' )
	f_{\omega' n' l' m' } (v, \vartheta, \varphi)
\label{eq16}  \eeq
which carry angular quantum numbers $( \Theta' , \Phi' )$.
The usual angular coordinates are
denoted by $\vartheta$ and $\varphi$. Using the
completeness relation \cite{Messiah}
of the spherical harmonics it is easy to show that
\beq  f_{\omega' n' \Theta' \Phi'} (v, \vartheta, \varphi)
\sim { \delta (\vartheta - \Theta' ) \,
	\delta ( \varphi - \Phi' ) \/ \sin ( \vartheta ) }
	 \label{eq17}  \eeq
showing that the angular wave packet modes are entirely
concentrated in the direction
$( \Theta', \Phi')$ indicated by their quantum numbers.
In the same manner we define angular wave packet modes at
${\cal J}^+$:
\beq  p_{\omega n \Theta \Phi} (u, \vartheta, \varphi) =
\sum_{l m} Y^\ast_{l m} (\Theta, \Phi )
	p_{\omega nlm} (u, \vartheta, \varphi)
\label{eq18}  \eeq
which are also concentrated in the direction of their
quantum numbers $(\Theta, \phi)$.

When the quantization at ${\cal J}^-$ and ${\cal J}^+$ is
performed with respect to these
bases, the creation and annihilation operators refer to the
angular wave packet modes
(\ref{eq16}) and (\ref{eq18}). The corresponding Bogoliubov
 transformation can be derived
easily from (\ref{eq8}). Using some well-known properties
of the spherical harmonics \cite{Messiah}
we find
\beq a^{out}_{\omega n \Theta \Phi} = \alpha^\ast_{\omega
\Theta\Phi\Theta'\Phi'}
b^{in}_{\omega'_\alpha n'_\alpha \Theta' \Phi'}
- \beta^\ast_{\omega \Theta\Phi \Theta'\Phi'}
b^{in\dagger}_{\omega'_\beta n'_\beta \Theta' \Phi'}
\label{eq19}\eeq
where the Bogoliubov parameters are now given by
 \beq\left. {\alpha_{\omega \Theta\Phi\Theta'\Phi'}
 \atop \beta_{\omega \Theta\Phi\Theta'\Phi'}}
 \right\} =
\mp i \left( {\kappa \/2\pi \omega} \right)^{1\/2}
e^{\pm \omega \pi / 2 \kappa} \,
\left| \Gamma\left( 1 + i {\omega \/\kappa} \right)\right|
	{\delta (\Theta - (\pi - \Theta')) \,
	\delta( \Phi - ( \Phi' + \pi)) \/ \sin \Theta} .
 \label{eq20}  \eeq
Their angular part shows that the wave packet simply
crosses the spatial origin of coordiates.
The radial part remains unchanged. The formulas
(\ref{eq13}) and (\ref{eq15}) are easily transcribed
in terms of angular wave packets. We now see that ingoing
particles in a mode $( \omega'_\beta,
n'_\beta, \Theta', \Phi')$ that travels behind the horizon
can influence the probability
for finding particles in the mode $(\omega, n, \pi -
\Theta', \Phi' - \pi)$ at ${\cal J}^+$.
Since they are spacelike separated this shows the
nonlocality of the stimulated emission process.
\vskip 0.5cm

\noindent
{\bf Acknowledgments}\\
We would like to thank J. Audretsch for helpful discussions.
This work was supported by the Commission of the
European Community DG XII. C.\ O.\ L.\ was also
supported by the Alexander von Humboldt Foundation.

\vskip 2 cm
\goodbreak
{\bf  Figure captions:} \hfill\break
\smallskip

Fig.\ 1. Penrose diagram for a Schwarzschild
black hole. Radially ingoing (outgoing) null geodesics are
characterized by $v= \hbox{const}$ ($u = \hbox{const}$).
$v_0$ denotes the null ray that reaches the event horizon $H^+$
at the moment of its formation. The ingoing wave packet
$( \omega'_\alpha, n'_\alpha, l, m)$ is the equivalent mode
to the outgoing wave packet $(\omega, n, l, m)$. Its mirror
mode $( \omega'_\beta, n'_\beta, l, -m)$, which propagates
behind the horizon, is obtained by `reflection' at the line
$v_0$.
\bigskip

Fig.\ 2. The information $I_{\omega n l m} = (S^{th}_{\omega nlm}
- S^{bh}_{\omega nlm})/ S^{th}_{\omega nlm}$ contained in
an outgoing mode $(\omega, n, l,m)$ about the initial wave
packet $( \omega'_\beta, n'_\beta, l, -m)$ that crosses the
horizon plotted as a function of
$|\beta_{\omega l m}|^2 = (\exp(\omega / T) -1)^{-1}$.
The number of ingoing particles is $r= 10, 100, 1000$
(bottom to top).


\begin{thebibliography}{99}
\bibitem{Zurek83} W. H. Zurek, in: P. Meystre,
M. O. Scully (eds): {\it Quantum Optics,
Experimental Gravity, and Measurement Theory},
Plenum Press, New York, 1983, p.87.
\bibitem{Barnett89} S. M. Barnett and S. J. D. Phoenix,
Phys. Rev. A {\bf 40}, 2404 (1989), {\it ibid.},
{\bf 44}, 535 (1991).
\bibitem{Preskill92} J. Preskill, Caltech Report CALT-68-1819,
hep-th 9209058 (1992).
\bibitem{Page93} D. N. Page, Preprint hep-th 9305040
(1993), to be published in the Proceedings of the 5th
Canadian Conference on General Relativity and Relativistic
Astrophysics 1993.
\bibitem{Page92} D. N. Page, in: V. De Sabbata and Z. Zhang
(eds): {\it Black Hole Physics}
	Kluwer Academic Publishers (1992), p. 185.
\bibitem{Wald92} R. M. Wald, {\it ibid.}, p. 55.
\bibitem{tHooft92} G. 't Hooft, {\it ibid.}, p. 381,
Nucl. Phys. B {\bf 355}, 138 (1990).
\bibitem{Giddings92} S. B. Giddings, University of California,
Santa Barbara Report UCSB-TH-92-36, hep-th 9209113 (1992).
\bibitem{Wald84} R. M. Wald, in: S. M. Christensen (ed.):
 {\it Quantum Theory of Gravity}, Adam Hilger, Bristol
(1984),p. 160.
\bibitem{Aharonov87} Y. Aharonov, A. Casher, and
S. Nussinov, Phys. Lett.. B {\bf 191}, 51 (1987).
\bibitem{Hawking76} S. W. Hawking, Phys. Rev. D
{\bf 14}, 2460 (1976).
\bibitem{Banks84} T. Banks, M. E. Peskin, and
L. Susskind, Nucl. Phys. B {\bf 244}, 125 (1984).
\bibitem{Page80} D. N. Page, Phys. Rev. Lett.
{\bf 44}, 301 (1980).
\bibitem{Wald86} R. M. Wald, Found. Phys. {\bf 16}, 501
(1986).
\bibitem{Audretsch92} J. Audretsch and R. M\"uller,
Phys. Rev. D {\bf 45}, 513 (1992).
\bibitem{Bekenstein93} J. D. Bekenstein, Phys. Rev. Lett.
{\bf 70}, 3680 (1993).
\bibitem{Schiffer93} M. Schiffer, Phys. Rev. D {\bf 48}, 1652
(1993).
\bibitem{Misner73} C. W. Misner, K. S. Thorne, and
J. A. Wheeler: {\it Gravitation}, W. H. Freeman,
New York (1973).
\bibitem{Hawking75} S. W. Hawking, Commun. Math. Phys.
{\bf 43}, 199 (1975).
\bibitem{Wald75} R. M. Wald, Commun. Math. Phys.
{\bf 45}, 9 (1975).
\bibitem{Audretsch93} J. Audretsch and R. M\"uller,
University of Konstanz Preprint (1993).
\bibitem{Gasperini90} M. Gasperini, Prog. Theor. Phys.
{\bf 84}, 899 (1990).
\bibitem{Abramowitz} M. Abramowitz and  I. A. Stegun
(eds.), {\it Handbook of Mathematical Functions}, Dover,
New York (1972), Eq. (6.1.50).
\bibitem{Yamamoto86} Y. Yamamoto and H. A. Haus,
Rev. Mod. Phys. {\bf 58}, 1001 (1986).
\bibitem{Araki70} H. Araki and E. Lieb, Commun. Math.
Phys. {\bf 18}, 160 (1970).
\bibitem{Parker75} L. Parker, Phys. Rev. D {\bf 12},
1519 (1975).
\bibitem{Parker71} L. Parker, Phys. Rev. D {\bf 3},
346 (1971).
\bibitem{Audretsch86} J. Audretsch and P. Spangehl,
Phys. Rev D {\bf 33}, 997 (1986).
\bibitem{Wald76} R. M. Wald, Phys. Rev. D {\bf 13}, 3176 (1976).
\bibitem{Wilczek93} F. Wilczek, Princeton Report IASSNS-HEP-93/12,
 hep-th/9302096 (1993).
\bibitem{Kiefer91} C. Kiefer and T. P. Singh, Phys. Rev.
D {\bf 44}, 1067 (1991).
\bibitem{Kiefer93} C. Kiefer, R. M\"uller and T. P. Singh,
University of Z\"urich Preprint ZU-TH 25/93, gr-qc 9308024.
\bibitem{Messiah} A. Messiah: {\it Quantum Mechanics}
Vol. 1, North Holland, Amsterdam (1970), Eqs. (B. 88),
(B. 91), (B. 92).
\end{thebibliography}
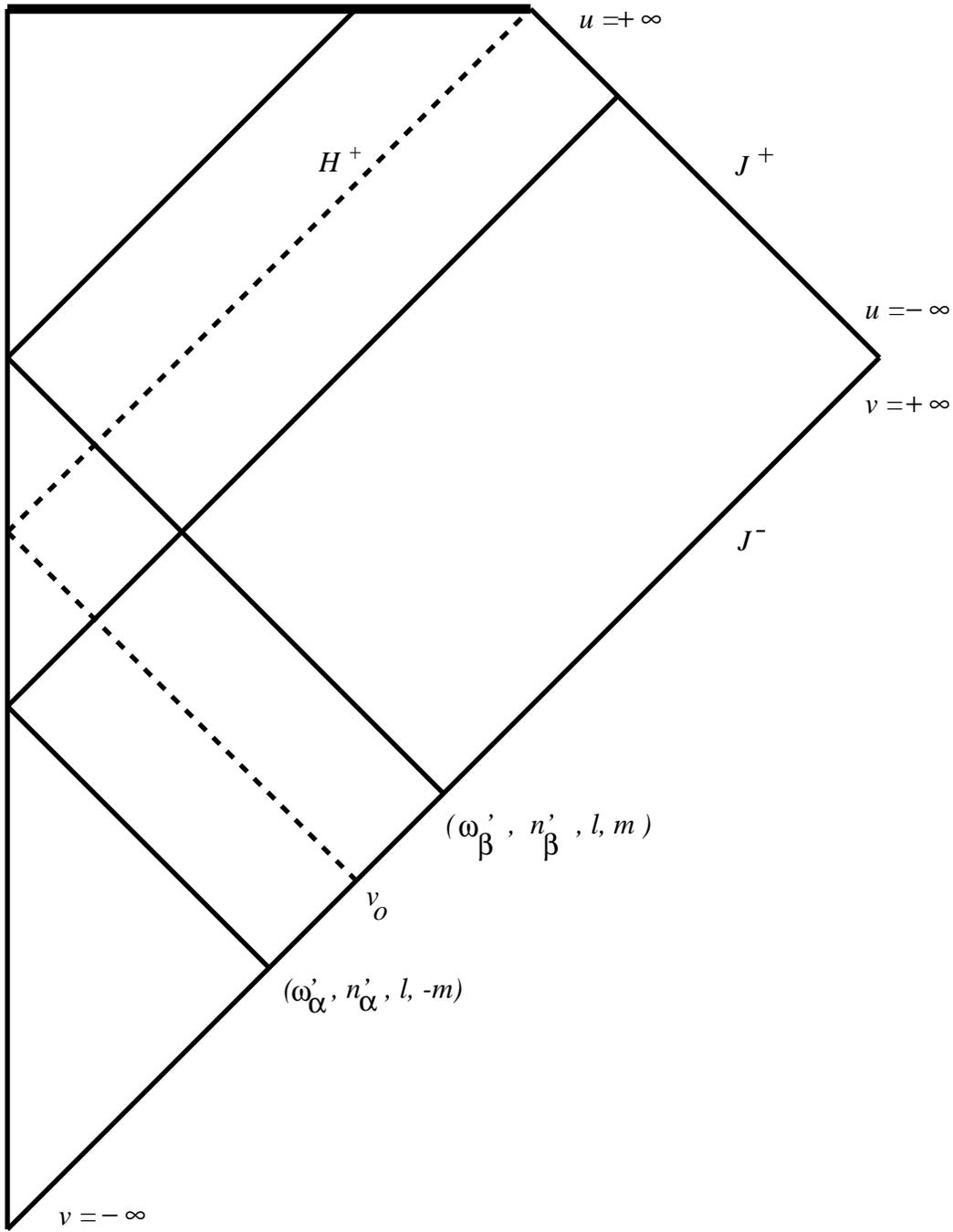
\end{document}